# Exciting surface plasmon polaritons in the Kretschmann configuration by light beam

A. P. Vinogradov,[1,2,3] A. V. Dorofeenko,[1,2,3] A. A. Pukhov,[1,2,3] and A. A. Lisyansky[4,5,*]

[1]*Institute for Theoretical and Applied Electromagnetics of Russian Academy of Sciences, 13 Izhorskaya, Moscow 125412, Russia*

[2]*Dukhov Research Institute of Automatics, 22 Suschevskaya, Moscow 127055, Russia*

[3]*Moscow Institute of Physics and Technology, 9 Institutskiy pereulok, Dolgoprudniy 141700, Moscow region, Russia*

[4]*Department of Physics, Queens College of the City University of New York, Queens, NY 11367, USA*

[5]*The Graduate Center of the City University of New York, New York, NY 10016*

We consider exciting surface plasmon polaritons in the Kretschmann configuration. Contrary to common belief, we show that a plane wave incident at an angle greater than the angle of total internal reflection does not excite surface plasmon polaritons. These excitations do arise, however, if the incident light forms a narrow beam composed of an infinite number of plane waves. The surface plasmon polariton is formed at the geometrical edge of the beam as a result of interference of reflected plane waves.

## I. INTRODUCTION

The Kretschmann configuration (KC) is one of the first setups, with which surface plasmon polariton (SPP) resonances were observed [1-6]. This configuration along with the Otto configuration [7] was used for measuring dielectric permittivities [8-14]. In the KC, a metal film and then the material under investigation (analyte) are deposited on the base of a dielectric prism. A beam of *p*-polarized light is transmitted through the prism (Fig. 1). Since the dielectric permittivity of the prism, $\varepsilon_p$, is greater than that of the analyte, there is a critical angle at which total internal reflection occurs. At some angle greater than the angle of total internal reflection, a sharp minimum is observed in the reflection coefficient, due to losses in the metal. This phenomenon is commonly referred to as attenuated total reflection (ATR) [2]. The tangential component of the wave vector, $k_x$, of the incident light corresponding to the ATR is close to the wavenumber of the SPP that may travel along the metal surface. Even though these wavenumbers are somewhat different, the possibility of resonant coupling of the incident wave with an SPP is customarily postulated, and excitation of the SPP is assumed to be the cause of the ATR [2,3].

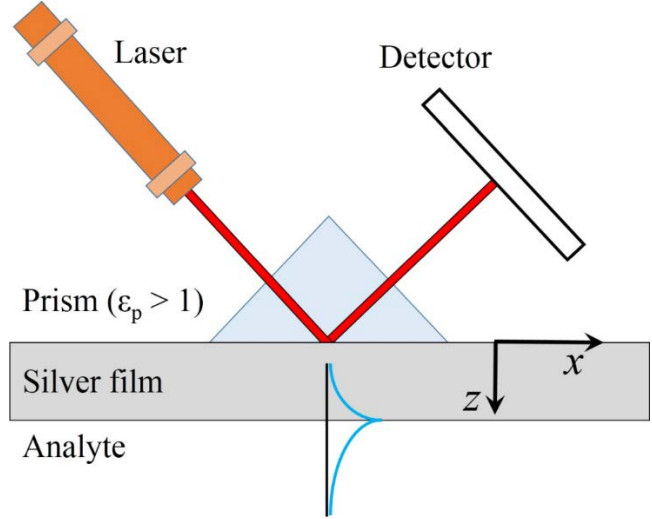

FIG. 1. The Kretschmann setup. Blue lines show profiles of resonantly excited fields.

It is well known that a plane wave propagating in a uniform medium cannot excite an SPP that travels along the interface between vacuum and metal. Indeed, the translational invariance along the interface requires the conservation of tangential components of the momentum, but the tangential component of the wave vector of the incident wave is smaller than that of an SPP [15]. Thus, an SPP can only be excited if the translational invariance of the system is violated. An example of such a violation is a local corrugation of the metal film. Then an SPP is part of the field scattered on this corrugation [3,4,13,16-19].

The reflection of a plane wave in the KC has a resonant nature. Due to the resonance, the reflection coefficient $r(k_x)$ has a pole for $k_x = k_{SPP}$ [3]. It seems that due to this fact it is usually presumed that in the KC, an SPP can be excited by a plane wave [1-3,7,15,20,21]. This is often justified by the fact that in the KC, the wavenumbers of the incident wave and the SPP coincide [3,15] (see also [2]) providing the possibility of the momentum conservation.

In the KC, the SPP is a leaky wave (see Ref. [15] and Section II); the tangential and normal components of the SPP wave vector have imaginary parts even in a lossless system. Consequently, the amplitude of such an SPP should decrease in the direction of propagation and increase away from the interface. The latter feature distinguishes an SPP from a reflected plane wave observed in the KC. Therefore, we consider that an SPP is excited when a contribution to the scattered field, which amplitude increases away from the interface, arises. Since such an increase is not observed in the experiment, a plane wave does not excite the SPP, and the latter cannot be a cause for ATR. Though the resonant nature of $r(k_x)$ and matching real parts of tangential wavenumbers are *necessary* for the SPP excitation, these conditions are *not sufficient*. In Sec. III, we discuss the relationship between the ATR dip and SPPs in detail. We show that ATR arises not due to the resonance of the reflection coefficient but due to its zero at $k_x = k_{SPP}$.

In this paper, we consider the KC for an incident narrow light beam. We demonstrate that if the ATR condition is fulfilled, then just beyond the geometrical edge of the beam, interference of the reflected plane waves composing the beam results in the formation of an SPP. We also show that evanescent waves play no role in forming ATR.

## II. LEAKY SURFACE PLASMON POLARITONS AT DIELECTRIC–METAL–VACUUM SANDWICH

Let us consider the properties of an SPP propagating in a layered dielectric–metal–vacuum system. For simplicity, let us begin with a lossless system.

If the metal film is sandwiched by the vacuum, then two types of SPPs can arise. These are SPPs with symmetric and antisymmetric, with respect to the central plane of the film, distributions of the magnetic field (Figs. 2a,b) [2,22]. In SPPs of both types, the intensities of the field on opposite film surfaces are the same, and away from the film, the field decays exponentially [2]. In a lossless system, the tangential wavenumbers of these SPPs are real and are greater than the vacuum wavenumber $k_0$. In this geometry, an incident plane wave can excite neither of these SPPs.

In a dielectric–metal–vacuum system, the symmetries mentioned above are broken [13]. The symmetric SPP is deformed in a way that at the metal–vacuum interface, the field intensity is higher than at the metal-dielectric interface (Fig. 2c). Below, we refer to this SPP as to the pseudosymmetric (*pS*). The antisymmetric SPP becomes the pseudoantisymmetric (*pA*) SPP with the maximum field intensity at the metal-dielectric interface (Fig. 2d).

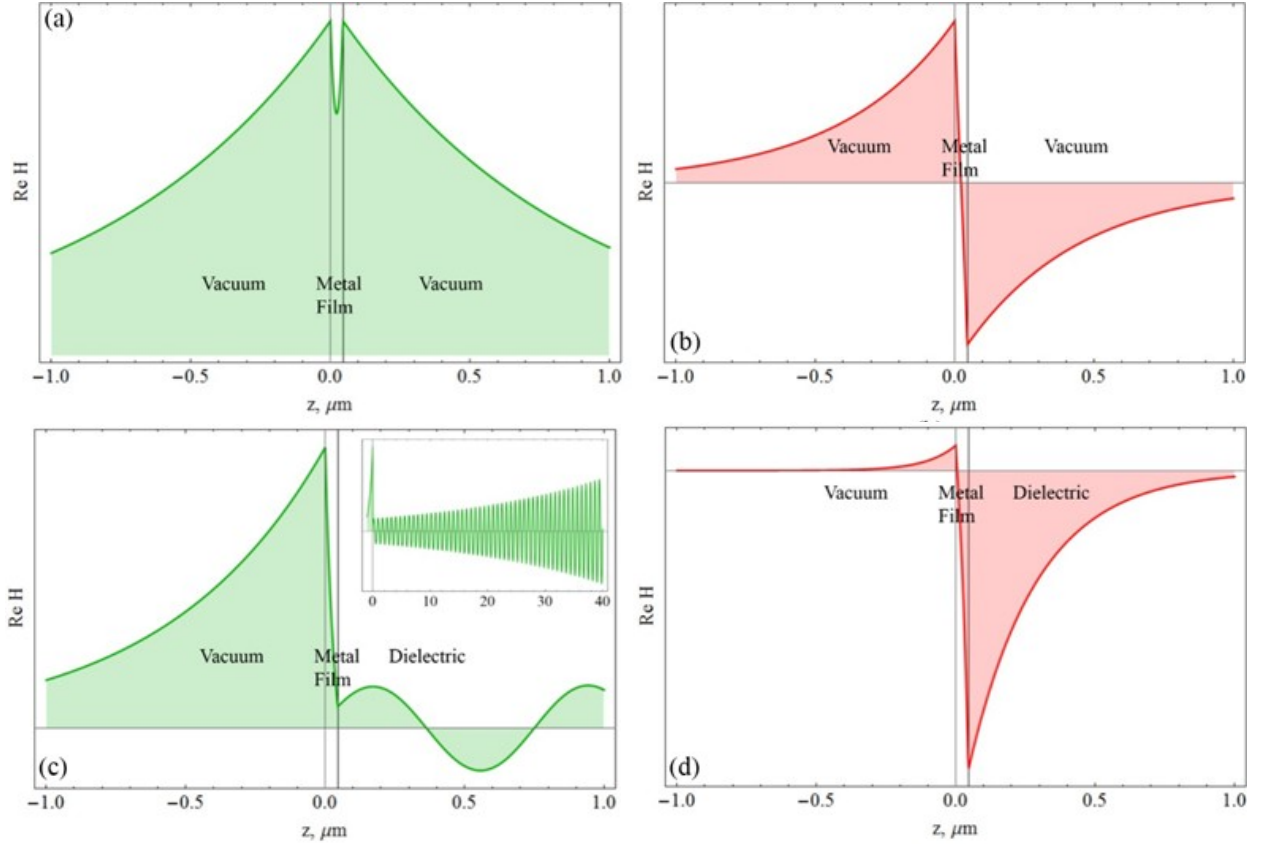

FIG. 2. Plasmonic solutions in (a, b) a symmetric (vacuum/metal/vacuum) system and (c, d) an asymmetric (vacuum/metal/dielectric) system. The inset shows the field amplitude increase in the dielectric away from the interface.

In a lossless case, even if the value of the permittivity $\varepsilon_D$ of dielectric differs significantly from unity, the *pA*-SPP remains a surface wave; its field decreases exponentially away from the film, and the tangential component of its wave vector, $k_x = k_{pA}$ is a real number, which is always greater than the wavenumber $k_\varepsilon = k_0\sqrt{\varepsilon_D}$ of a plane wave in the dielectric of the prism (Fig. 2d). If losses in the metal are taken into account, $k_{pA}$ becomes a complex number. The intensity of the field of the *pA*-SPP decays as the wave propagates along the film. The field is no longer evanescent, but it still decays away from the metal surface. Moreover, it transfers the energy into the metal film [2,3] from both sides of the film.

In contrast to the $pA$-SPP, the wavenumber $k_{pS}$ of the $pS$-SPP becomes complex even in the lossless case. The real part of $k_{pS}$ is smaller than the wavenumber of a plane wave propagating through the dielectric prism, $k_0 < \operatorname{Re} k_{pS} < k_0\sqrt{\varepsilon_D} = k_\varepsilon$. As a result, in the prism, an exponential decay of the SPP field changes to oscillations and the SPP becomes a leaky wave [23-25], which amplitude grows away from the interface (see inset in Fig. 2c). This wave loses energy by radiating it at an angle $\alpha = \sin^{-1}(\operatorname{Re} k_{SPP}/k_0)$ into the prism [15]. The normal component of the Poynting vector of this wave is directed outward from the vacuum toward the metal film and further into the prism. On the interface, simultaneously with the energy loss, the magnetic field of a leaky wave decreases in the direction of propagation. The magnetic fields $H(x_0, 0)$ at some point $x_0$ on the interface ($z=0$) is smaller than the field $H(x_0, z_0)$ at a distance $|z_0|$ above the interface, $H(x_0, z_0) > H(x_0, 0)$, i.e., the amplitude of the leaky wave increases in the direction of the normal component of the Poynting vector. Such a behavior occurs because $H(x_0, z_0)$ arises from the point $(x_0 - |z_0|\tan\alpha, 0)$. Due to damping of the leaky wave when it propagates along the interface, $H(x_0 - |z_0|\tan\alpha, 0) > H(x_0, 0)$. The behavior of this plasmon is similar to the behavior of the resonant eigenmode for 1-D dielectric-slab-resonator, from which field leaks with time. These two problems are related to one another via mapping $\{x, z\}_{KC} \Rightarrow \{t, z\}_{DSR}$. Since a dielectric-slab is an open resonator, it radiates. The radiated field is proportional to $\exp(-i\omega t + ikx)$. Even for a lossless dielectric, due to radiation, the amplitude of the field in the eigenmode inside the slab diminishes over time. Therefore, the eigenfrequency is a complex value with a negative imaginary part, and the amplitude of the radiated field emitted at the time moment $t_2$ is smaller than that emitted at a previous time moment $t_1 < t_2$. At a fixed time $t > t_1, t_2$ these radiated fields reach the points $z_1 = c(t - t_1)$ and $z_2 = c(t - t_2)$, respectively. Since $t_1 < t_2$, we have that $z_1 > z_2$. As it is mentioned above, the field in the point $z_1$ is larger than that in the point $z_2$. Thus, at a fixed time moment, the radiated field increases as going away from the slab. The same increase is characteristic of the leaky surface plasmon.

The situation with a leaky plasmon is analogous to a quantum mechanics problem of a slow particle scattering at a shallow well [26]. Treating the situation with a shallow well implies that though there is no real energy level, the scattering is resonant due to an existence of the virtual level. The pole that corresponds to this virtual level lies on a non-physical sheet of the Riemann surface. As a consequence, the wave function of the virtual state grows at the infinity similar to the field of a leaky plasmon.

For a gold film, frequency dependencies of real and imaginary parts of $k_{pS}$ are shown in Fig. 3. It is important that in lossy films, a $pS$-SPP is always a leaky wave whose dispersion curves lie above the light cone of the prism.

Thus, the main features of a $pS$-SPP that distinguishes it from a common plasmon and a reflected plane wave is that even in a lossless medium, the wavenumber is complex, because this plasmon is a leaky wave. Its field increases away from the dielectric-metal interface. We use this feature as the criterion for the $pS$-SPP excitation.

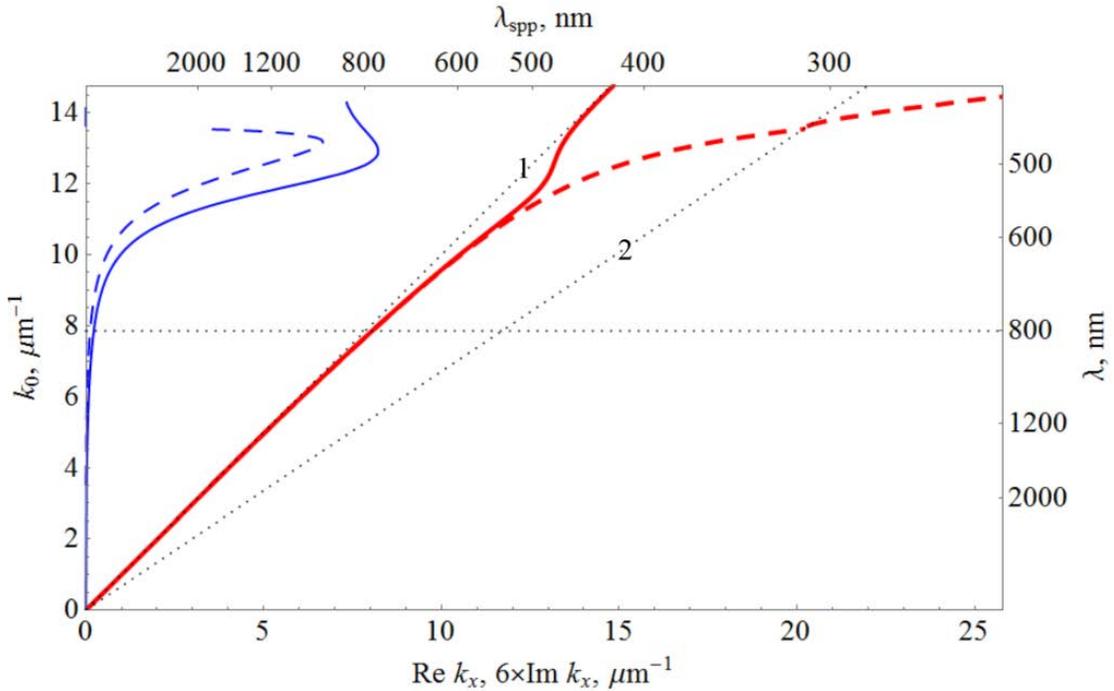

FIG. 3. Dispersions of real (red curves) and imaginary (blue curves) parts of the wavenumber of the $pS$-SPP in a vacuum/gold/dielectric system with (solid curves) and without (dashed curves) losses. The horizontal dotted line corresponds to the free space wavelength $\lambda = 800$ nm, which is used in further calculations, and inclined dotted lines show the light cones

of the vacuum (curve 1: $k_0 = k_x$) and dielectric (curve 2: $k_0 = k_x / \sqrt{\varepsilon_D}$). The metal (Au) thickness is $d_M = 45$ nm, and the dispersion is taken from Ref. [27]. The dielectric permittivity is $\varepsilon_D = 2.22$.

## III. THE KRETSCHMANN EFFECT

It is commonly believed that in the KC the phenomenon of attenuated total reflection (ATR) [2] is caused by an excitation of SPPs [2,3]. In this Section we show that this is not so.

Let us consider the reflection coefficient $r(k_x)$ in the KC as a function of the complex tangential wavenumber of an incident wave $k_x$,

$$r(k_x) = \frac{(\zeta_D - \zeta_M)(\zeta_V + \zeta_M)\exp(-ik_{zM}d) - (\zeta_D + \zeta_M)(\zeta_V - \zeta_M)\exp(ik_{zM}d)}{(\zeta_M + \zeta_D)(\zeta_V + \zeta_M)\exp(-ik_{zM}d) + (\zeta_M - \zeta_D)(\zeta_V - \zeta_M)\exp(ik_{zM}d)}, \quad (1)$$

where $d$ is the thickness of the metal layer, $k_{zV} = \sqrt{k_0^2 - k_x^2}$, $k_{zM} = \sqrt{\varepsilon_M k_0^2 - k_x^2}$, and $k_{zD} = \sqrt{\varepsilon_D k_0^2 - k_x^2}$ are normal to the surface components of the wave vectors in the vacuum, metal, and dielectric, respectively; $\zeta_V = k_{zV}/k_0$, $\zeta_M = k_{zM}/k_0\varepsilon_M$, and $\zeta_D = k_{zD}/k_0\varepsilon_D$ are the surface impedances in the same media.

The function $r(k_x)$ may have poles and branching points. The poles of the reflection coefficient, $r(k_x)$, arise due to zeros of the denominator in Eq. (1),

$$(\zeta_D + \zeta_M)(\zeta_V + \zeta_M)\exp(-ik_M d) + (\zeta_M - \zeta_D)(\zeta_V - \zeta_M)\exp(ik_M d) = 0. \quad (2)$$

Eq. (2) coincides with the dispersion equations of the *pA*- and *pS*-SPPs on the metal film [13,28]. As shown in the next Section, the pole at the point $k_x = k_{pS}$ corresponds to the *pS*-SPP that is a leaky wave, while the pole at $k_x = k_{pA}$ corresponds to the *pA*-SPP. Below we focus on *pS*-SPP.

The reflection coefficient, being a function of the impedance values, has branching points $k_x = \pm k_0$ and $k_x = \pm k_0\sqrt{\varepsilon_D}$ due the square roots $k_{zV} = \sqrt{k_0^2 - k_x^2}$ and $k_{zD} = \sqrt{k_0^2 \varepsilon_D - k_x^2}$. In order to find a single-valued branch of $r(k_x)$, we should make cuts on the complex plane and define impedances at some point. Usually one makes a cut along the negative part of the real axis of $k_x$. Such a cut provides an increase (decrease) of the amplitude of the plane wave in the direction of its phase velocity when this wave travels through a dissipative (active) medium. For $r(k_x)$, the corresponding cuts are shown in Fig. 4a. In this case, the pole of the function $r(k_x)$ is on the unphysical branch of the Riemann surface of the reflection coefficient (see Fig. 5a). The situation, in which a singular point (pole) of a function is on the unphysical Riemann surface sheet of $r(k_x)$ is well-known in quantum mechanics (resonant scattering from a quasidiscrete level [29] and in the theory of spontaneous emission in media with spectral peculiarities [30]).

Even though a pole of the function $r(k_x)$ is on the unphysical Riemann sheet, it strongly affects the behavior of the reflection coefficient on the real axis. Below, we demonstrate that the position of this pole along with the zero of the reflection coefficient affects ATR. For this purpose, it is convenient to deform cuts and to redefine the analytical branch of $r(k_x)$ so that the values of this function on the real axis remain the same, but the poles would be on a newly formed sheet of the Riemann surface. The new cuts for the square root function should be made in the direction along the negative imaginary axis. The corresponding cuts for $r(k_x)$ are shown in Fig. 4b. In this case, we can observe the poles (see Fig. 5b).

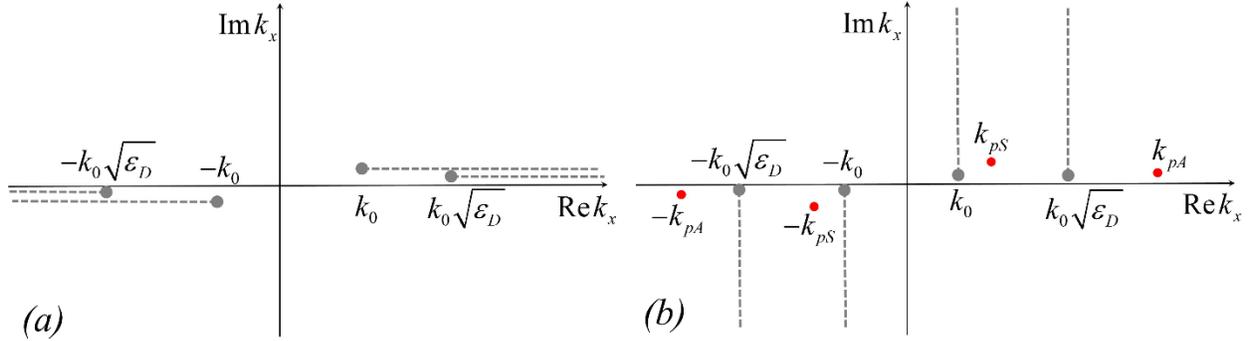

*(a)*        *(b)*

FIG. 4. Singularities of the reflection coefficient $r(k_x)$: poles that correspond to SPPs are shown by dots and cuts are shown by dashed lines. (*a*) and (*b*) show, respectively, conventional and modified cuts for the square root.

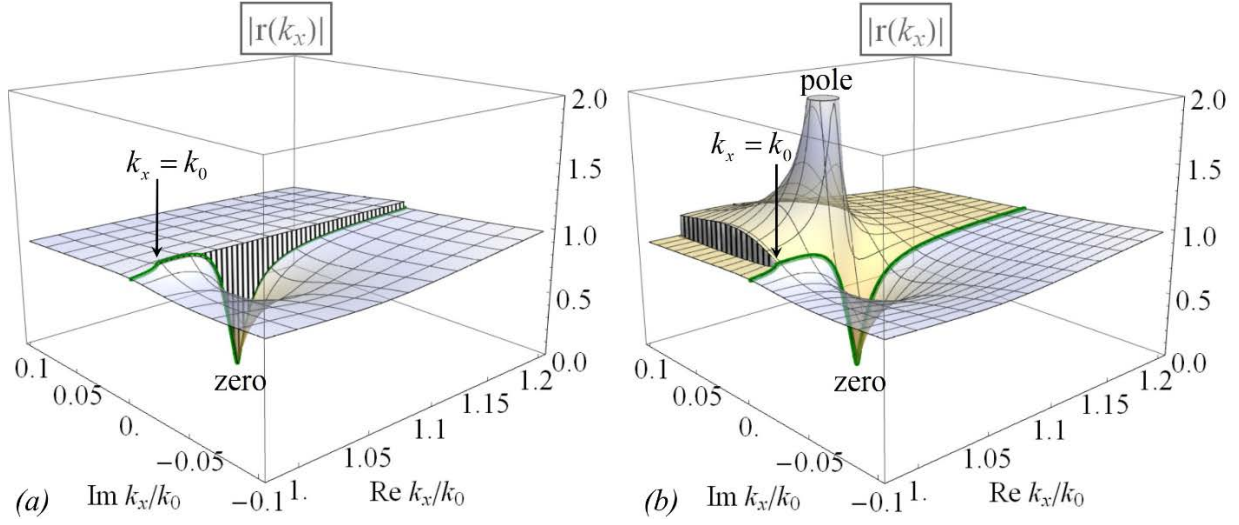

FIG. 5. The absolute value of the reflection coefficient $|r(k_x)|$ calculated for (a) standard and (b) modified cuts, respectively. The jump of $|r(k_x)|$ on the cut is shown by the vertical line shading. The solid thick green line corresponds to the cross-section of the $|r(k_x)|$-surface by the plane $\mathrm{Im}\,k_x = 0$.

Equation (1) can now be rewritten as [31]

$$r(k_x) = \frac{r_{DM} + r_{MV}\exp(-2\kappa)}{1 + r_{DM}r_{MV}\exp(-2\kappa)}, \qquad (3)$$

where the subscripts *D*, *M*, and *V* indicate that the corresponding quantity belongs to dielectric, metal, and vacuum, $r_{ab} = (\zeta_a - \zeta_b)/(\zeta_a + \zeta_b)$ is the reflection coefficient of the incident wave on interface between two half-spaces *a* and *b* ( *a* and *b* stand for *D*, *M*, or *V*), $\zeta_a = \sqrt{k_0^2 \varepsilon_a - k_x^2}/(k_0 \varepsilon_a)$ is a surface impedance of the *TM*-polarized wave, $\kappa = \sqrt{k_x^2 - k_0^2 \varepsilon_M}\,d_M$ is the complex-valued attenuation coefficient. The transmission and reflection coefficients for the amplitude, $t_{ab} = 2\zeta_a/(\zeta_a + \zeta_b)$ and $r_{ab}$, respectively, are connected via the relation $t_{ab} = 1 + r_{ab}$ that results from the continuation of the fields on the interface. This allows us to rewrite Eq. (1) as

$$r(k_x) = r_{DM} + \frac{t_{DM}r_{MV}t_{MD}\exp(-2\kappa)}{1 - r_{MV}r_{MD}\exp(-2\kappa)}. \qquad (4)$$

All coefficients $r_{ab}$ and $t_{ab}$ are analytical functions of the tangential wavenumber $k_x$; they vary slowly at the length-scale of the width of the plasmon resonance. A sharp resonance dependence of $r(k_x)$ arises due to the pole in Eq. (4). This pole corresponds to an SPP. It arises when the denominator in Eq. (4) is zero,

$$1 - r_{MV} r_{MD} \exp(-2\kappa) = 0. \tag{5}$$

One of the solutions of Eq. (5), $k_x = k_{pS}$, corresponds to the *pS*-SPP. Since even in lossless media, $k_{pS}$ is a complex number, there are no real solutions of Eq. (5). However, due to the smallness of the imaginary part of $k_{pS}$ and a slow dependence of $r(k_x)$ on $k_x$ near the pole, the denominator in Eq. (5) can be expanded into the Taylor series, $\left[1 - r_{MV} r_{MD} \exp(-2\kappa)\right] \sim (k_x - k_{pS})$. This allows us to reduce Eq. (5) to

$$r(k_x) \simeq r_{DM} + \frac{\alpha}{k_x - k_{pS}}, \tag{6}$$

where

$$\alpha = -\frac{t_{DM} r_{MV} t_{MD} \exp(-2\kappa)}{\partial \left[r_{MV} r_{MD} \exp(-2\kappa)\right]/\partial k_x}\bigg|_{k_x = k_{pS}} \tag{7}$$

is the residue of the pole at $k_x = k_{SPP}$. This result is a generalization of well-known Eq. (2.18) in Ref. [3] (see also [15]). Equation (6) differs from Eq. (2.18) in Ref. [3] in that the former is obtained for the *amplitude* reflection coefficient, whereas the latter concerns the *intensity* reflection coefficient. Therefore, Eq. (6) in addition to the amplitude takes into account the phase of the reflected signal.

Equation (6) shows that the reflected wave can be considered as a result of interference of two waves. The first term of this equation, $r_{DM}$, is non-resonant, while the second is resonant. Such a response is known as the Fano resonance that is characterized by asymmetric line shape with a minimum [32,33]. To emphasize this fact, it is convenient to rewrite Eq. (6) as

$$r(k_x) \approx r_{DM} \frac{k_x - k_{zero}}{k_x - k_{SPP}}, \tag{8}$$

where $k_{zero} = k_{SPP} - \alpha / r_{DM}$ is the position of the zero of the reflection coefficient. In a lossless medium, $k_{zero}$ and $k_{SPP}$ are mutually complex conjugated (see Fig. 6a and the points labeled by "a" in Fig. 7). Generally, this is not the case. If the zero is on the real axis, then it leads to a maximum of ATR (see Ref. [3]).

Let us go back to Eq. (1) to consider how losses in metal affect the position of the zero of the reflection coefficient. When losses are absent, due to the symmetry in the positions of the pole and the zero with respect to the real axis, one observes unitary values $|r(k_x)|=1$ at the real axis (solid line in Fig. 6a). Losses result in shifting $k_{zero}$ toward the real axis (open circle labeled by "b" in Fig. 7), and $k_{SPP}$ shifts away from the real axis. As a result, the compensation due to symmetry in the positions of the pole and the zero does not happen, and the reflection coefficient decreases (Fig. 6b). The value of $k_x$ at which the zero crosses the real axis corresponds to the equality of Joule and radiation losses. A further increase in losses moves $k_{zero}$ into the upper half-plane (open circle labeled by "c" in Fig. 7) resulting in smoothing of the minimum (see Fig. 6c). Similar behavior of the minimum of the reflection coefficient is observed in experiment [5].

Finally, let us consider a hypothetical case of an active metal that amplifies the field (circles labeled by "d" in Fig. 7). One can see that the pole-zero couple moves down in Fig. 7. In this case, instead of the minimum, a maximum in the dependence $|r(k_x)|$ at real axis arises (see also [14]). At the amplification that exactly compensates for radiation losses, $\text{Im}\,k_{pS}$ becomes zero, the leaky SPP becomes a plane wave, and the maximum of the reflection coefficient tends to infinity. It indicates the beginning of lasing.

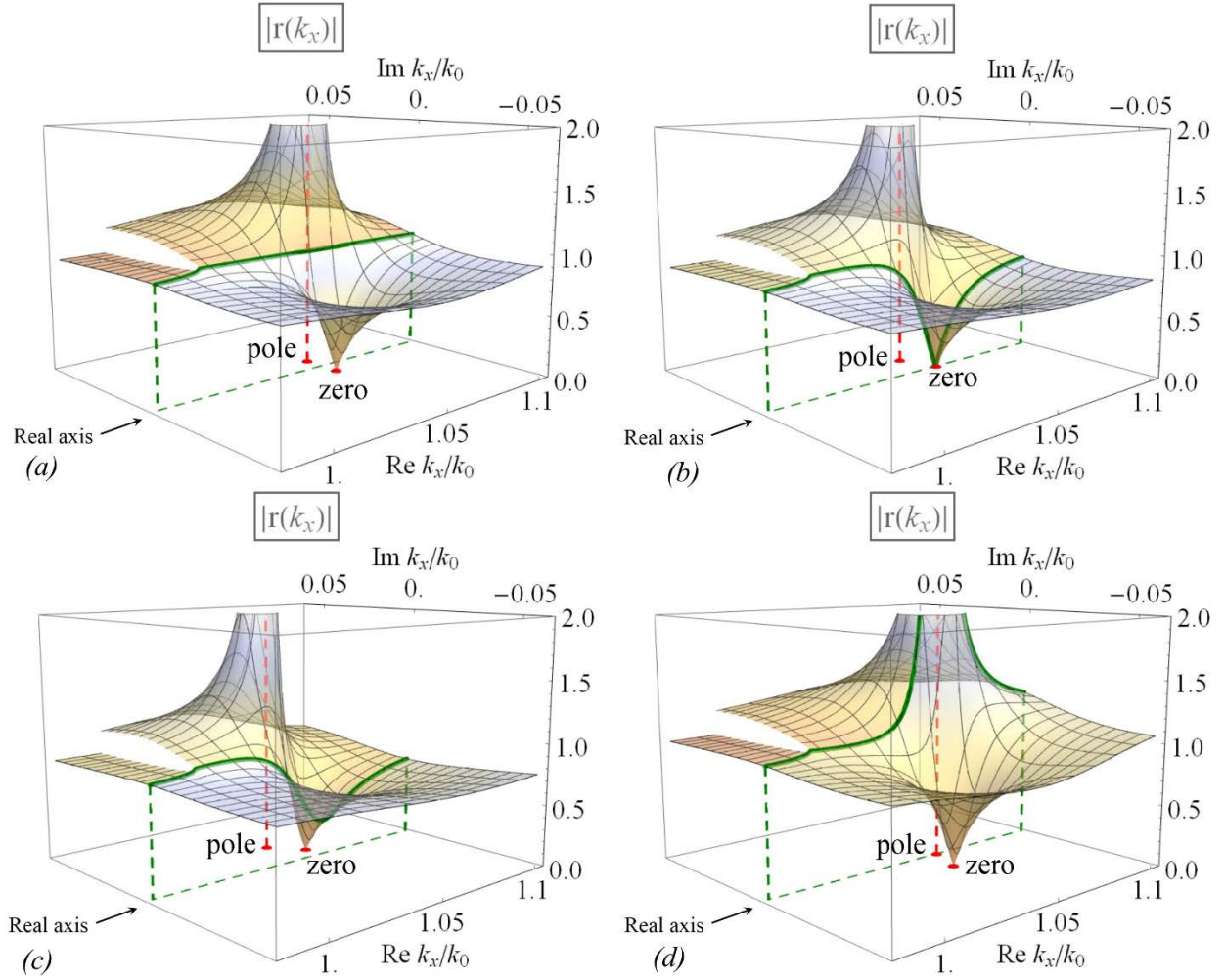

FIG. 6. Changes of the absolute value of the reflection coefficient $|r(k_x)|$ at different levels of loss in a metal. The metal permittivity is taken in the form $\varepsilon = \varepsilon' + i\gamma\varepsilon''$. (*a*) A hypothetical lossless case, $\gamma = 0$ ($\varepsilon_M = -14.3$ at $\lambda = 630$ nm); (*b*) the case of real losses in silver, $\gamma = 1$ ($\varepsilon_M = -14.3 + 1.18i$); (*c*) a hypothetical system with loss twice as large as that in silver, $\gamma = 2$ ($\varepsilon_M = -14.3 + 2.36i$); (*d*) a hypothetical system with an active metal, $\gamma = -1$ ($\varepsilon_M = -14.3 - 1.18i$). The solid thick green line denotes the cross section of the surface $|r(k_x)|$ with the plane $\operatorname{Im} k_x = 0$.

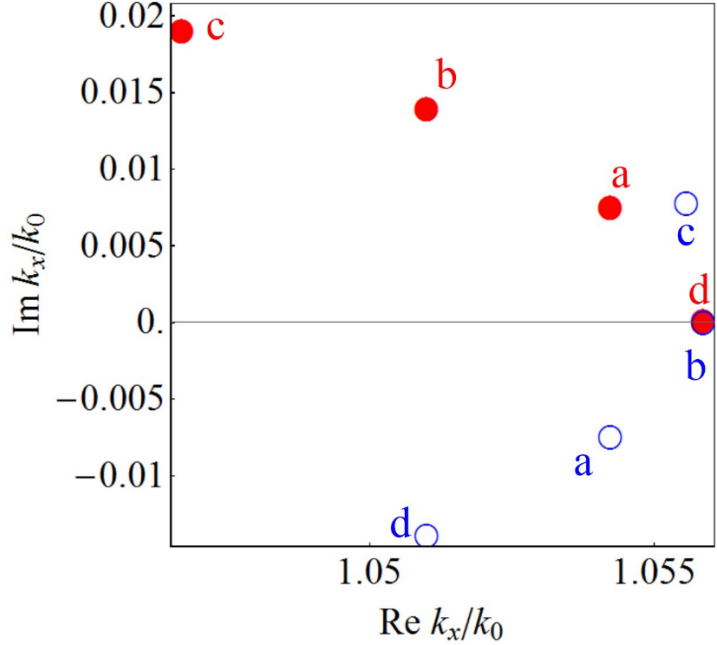

FIG. 7. The positions of $k_{pS}$ (solid red circles) and $k_{zero}$ (open blue circles) for different losses. Letters correspond to the cases shown in Fig. 6.

To conclude, the ATR arises due to the Fano resonance when $k_{zero}$ reaches the real axis. An excitation of an SPP ($k_{pS}$ crosses the real axis) is possible in the hypothetical gain system. In this case, instead of ATR, the lasing occurs.

## IV. EXCITATION OF A SURFACE PLASMON POLARITON IN THE KRETSCHMANN CONFIGURATION

As discussed in Introduction, to excite SPPs, translational invariance of a system should be broken by inhomogeneity of the surface or of an incident wave. Below, we consider an excitation of an SPP by a bounded beam. When the beam angle of incidence is near the angle corresponding to ATR, an additional field that extends far beyond the geometrical boundary of the reflected beam arises. In the Kretschmann geometry, this field has been observed in computer simulations [34] and experiments [35]. This effect has been interpreted as the excitation of an SPP. Let us consider this phenomenon in detail.

We assume that the beam is formed by a plane wave with $k_x = k_x^{inc}$ passing through a slit $X_0 - a/2 \le x \le X_0 + a/2$ that is situated in the plane $z = Z_0 < 0$ (Fig. 8a). The slit position $X_0 = Z_0 k_x^{inc} / \sqrt{k_0^2 \varepsilon_D - (k_x^{inc})^2}$ is chosen so that at the metal surface, $z = 0$, the center of the beam is at the point $x = 0$. The distance $|Z_0|$ between the slit and the metal surface is taken large enough ($10\lambda$) to assure that an SPP is not excited by the near field of the slit. In the slit plane, the magnetic field $H_i(x, Z_0)$ is equal to $H_0 \exp(ik_x^{inc}(x - X_0))$ inside and to zero outside the slit (the upper part of Fig. 8b). This field may be presented as a sum of plane waves:

$$H_i(x, Z_0) = \int_{-\infty}^{\infty} h_i(k_x, Z_0) \exp(ik_x x) dk_x, \tag{9}$$

where the spatial spectrum of the field just beyond the screen,

$$h_i(k_x, Z_0) = \frac{1}{2\pi} \int_{-\infty}^{\infty} H_i(x, Z_0) \exp(-ik_x x) dx$$
$$= \frac{H_0}{2\pi} \int_{X_0 - a/2}^{X_0 + a/2} \exp(ik_x^{inc}(x - X_0)) \exp(-ik_x x) dx, \tag{10}$$

evaluates to

$$h_i(k_x, Z_0) = H_0 \frac{\sin[(k_x - k_x^{inc})a/2]}{\pi(k_x - k_x^{inc})} \exp(-ik_x X_0). \tag{11}$$

By using the dispersion relation for the plane wave in vacuum, it is possible to write the expression for the incident field at some plane $z > Z_0 = \text{const}$,

$$H_i(x, z) = \int_{-\infty}^{\infty} h_i(k_x, z) \exp(ik_x x) dk_x, \tag{12}$$

with

$$h_i(k_x, z) = h_i(k_x, Z_0) \exp\left[i\sqrt{k_0^2 \varepsilon_D - k_x^2}(z - Z_0)\right]. \tag{13}$$

In particular, at the prism/metal interface ($z = 0$) we obtain:

$$h_i(k_x, 0) = H_0 \frac{\sin[(k_x - k_x^{inc})a/2]}{\pi(k_x - k_x^{inc})} \exp\left(-ik_x X_0 - i\sqrt{k_0^2 \varepsilon_D - k_x^2} Z_0\right). \tag{14}$$

The incident beam at the same interface,

$$H_i(x,0) = \int_{-\infty}^{\infty} h_i(k_x,0)\exp(ik_x x)\,dk_x, \qquad (15)$$

is evaluated numerically (the lower part of Fig. 8b). As one can see, even though at the prism-metal interface, the field intensity distribution $|H_i(x,0)|^2$ is similar to the initial distribution $|H_i(x,Z_0)|^2$, the beam shape is markedly blurred by diffraction. This effect is more obvious at larger scale (see the solid curve in Fig. 12a).

Let us note that the distance $|Z_0|$, which equals $10\lambda$ ($\lambda = 2\pi/k_0$) in our calculations, is large enough for evanescent waves ($k_x > k_0$) to decay. This is confirmed by an additional calculation, in which the amplitudes of all the evanescent waves are manually set to zero. In this calculation, the fields obtained are not noticeably different from these obtained with the full calculation in which the evanescent waves are taken into account.

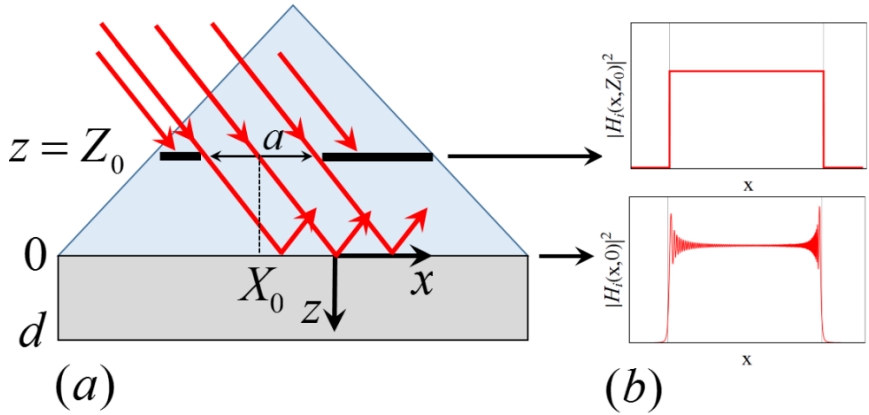

FIG. 8. (*a*) The geometry of the system for observing an excitation of an SPP at the edge of a beam. (*b*) Magnetic field intensity profiles in the *incident* wave at the slit level, $z = Z_0 < 0$, (the top figure) and at the metal film, $z = 0$, (the bottom figure).

By representing the total field of a plane wave, we reduce the problem of an incidence of a bounded beam to the problem of an incidence of a plane wave that has a well-known solution for the magnetic field $g(k_x, z)$:

$$g(k_x, z) = \begin{cases} \exp(ik_{zD}z) + r(k_x)\exp(-ik_{zD}z), & z < 0, \\ a(k_x)\exp(ik_{zM}z) + b(k_x)\exp(-ik_{zM}z), & 0 < z < d, \\ t(k_x)\exp(ik_{zV}(z-d)), & z > d. \end{cases} \qquad (16)$$

where, the complex amplitudes $a$, $b$, $t$, and $r$ can be found by matching solutions on the boundaries [31,36,37]. In particular, the amplitude of the reflected wave is given by Eq. (1). The total field of the beam can be found from Eqs. (14) and (16) by using to the following relation,

$$H(x, z) = \int_{-\infty}^{\infty} h_i(k_x, 0) g(k_x, z) \exp(ik_x x) dk_x. \qquad (17)$$

In particular, the reflected field $H_r$ is

$$H_r(x, 0) = \int_{-\infty}^{\infty} h_i(k_x, 0) r(k_x) \exp(ik_x x) dk_x. \qquad (18)$$

We consider a scattered part of the magnetic field, which is defined as the total field without the incident field:

$$g_S(k_x, z) = \begin{cases} r(k_x)\exp(-ik_{zD}z), & z < 0, \\ a(k_x)\exp(ik_{zM}z) + b(k_x)\exp(-ik_{zM}z), & 0 < z < d, \\ t(k_x)\exp(ik_{zV}(z-d)), & z > d, \end{cases} \qquad (19)$$

and

$$H_S(x, z) = \int_{-\infty}^{\infty} h_i(k_x, 0) g_S(k_x, z) \exp(ik_x x) dk_x. \qquad (20)$$

The absolute value of the intensity of the scattered magnetic field $H_S(x, z)$ on the *xz*-plane calculated with the use of Eq. (20) is shown in Fig. 9. Recall that the angle of incidence of the beam is chosen in the way that it would provide the minimum in the reflection coefficient. In this case, the reflected field is very weak inside the geometric optics boundaries of the beam (see the area between the dashed lines in Fig. 9) except for the regions near the geometrical boundaries of the beam.

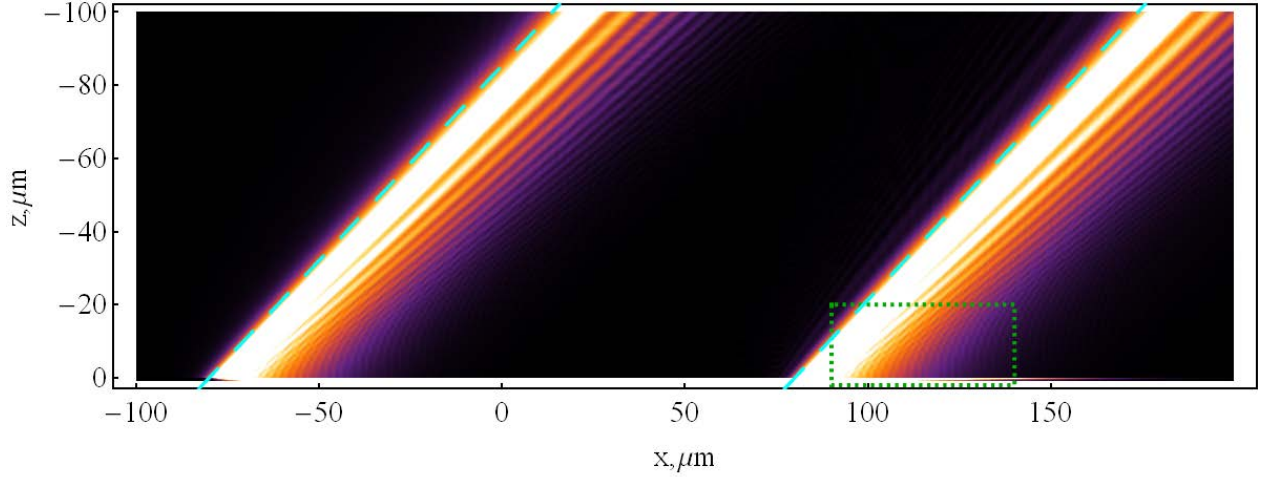

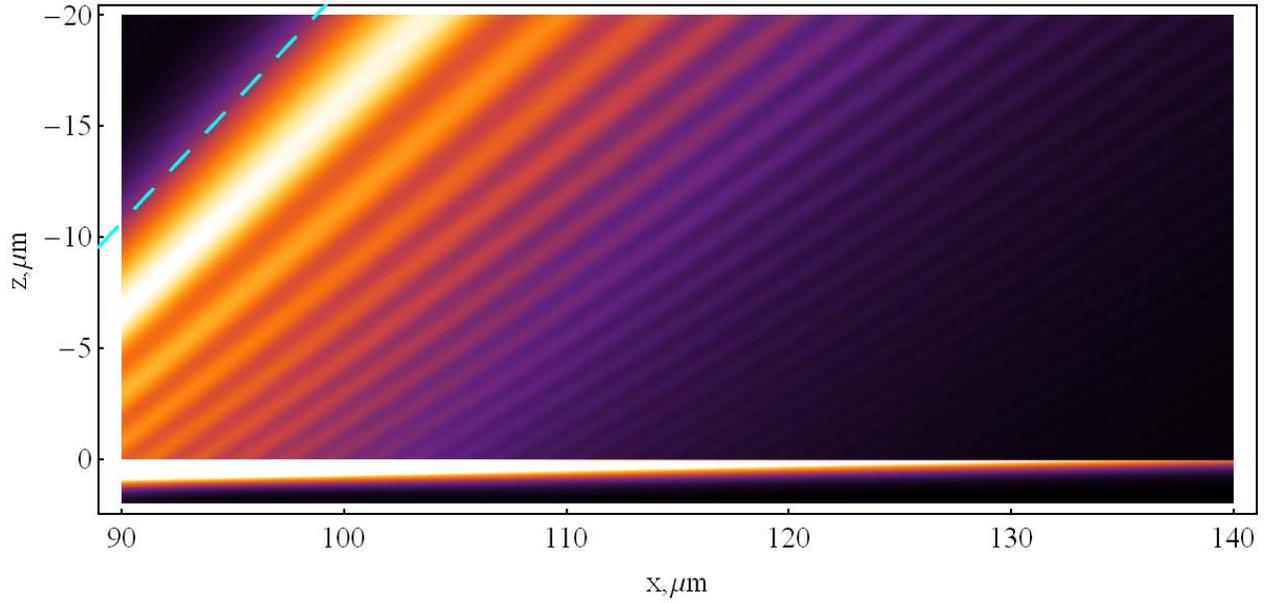

FIG. 9. The distribution of the intensity of the scattered part of the magnetic field of the light beam ($\lambda = 0.8\,\mu\text{m}$) in the Kretschmann effect: (a) a general view and (b) a magnified view of the dotted rectangle in Fig. 9a. The regions $z<0$, $0<z<0.045$ μm, and $z>0.045$ μm are filled by the dielectric, metal, and vacuum, respectively. The dashed lines show the geometrical boundaries of the beam.

Inside the geometric boundaries of the beam, $|x|<a/2$, due to symmetric properties of the integrand, integral (20) is very small. This becomes evident if we consider the point $x=0$. Note, that the spatial spectrum of the field $h_i(k_x,0)$ is a symmetric function with respect to $k_x = k_x^{inc}$ (see Fig. 10), while the reflection coefficient is nearly antisymmetric (Fig. 11). Therefore, at the point $x=0$, the total field is very small. Since the integrand includes the factor $\exp(ik_x x)$, this condition changes and the field slightly increases as we move toward the boundaries of the beam (see the dashed line in Fig. 10).

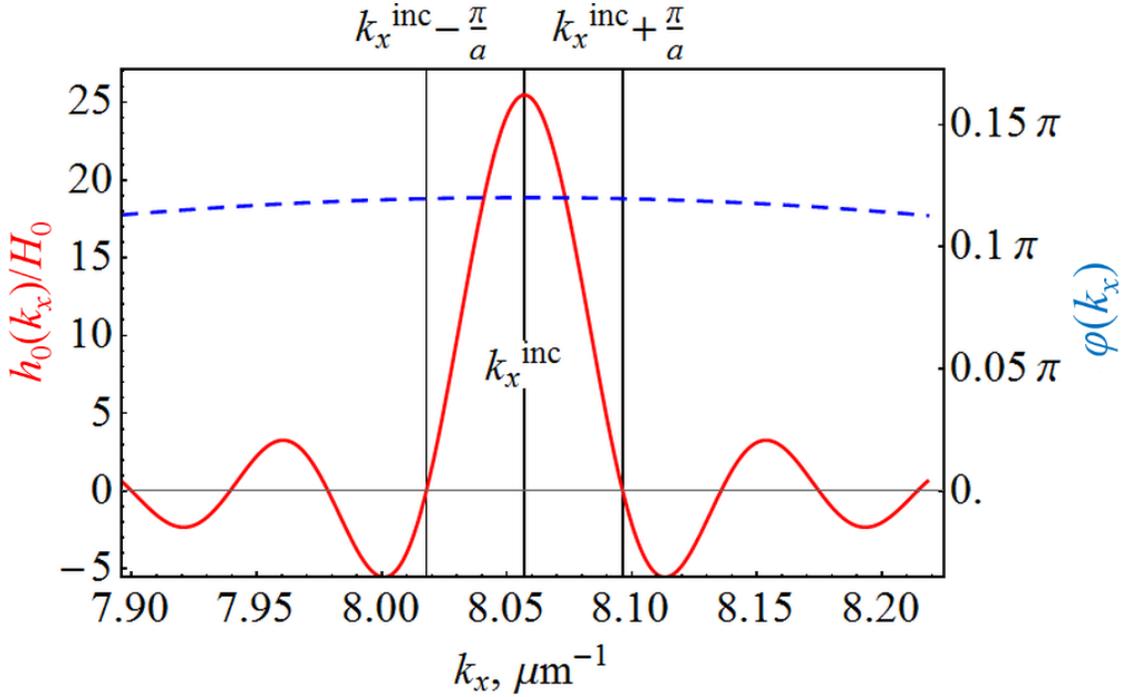

FIG. 10. The spatial spectrum of the wave at the dielectric-metal interface in terms of the functions $h_0(k_x)$ (the solid line) and $\varphi(k_x)$ (the dashed line). These functions are introduced as $h_0(k_x) = H_0 \sin\left[(k_x - k_x^{inc})a/2\right]/\left[\pi(k_x - k_x^{inc})\right]$ and $\varphi(k_x,0) = -k_x X_0 - \sqrt{k_0^2 \varepsilon_D - k_x^2} Z_0$, so that $h_i(k_x,0) = h_0(k_x)\exp\left[i\varphi(k_x)\right]$ [see Eq. (14)].

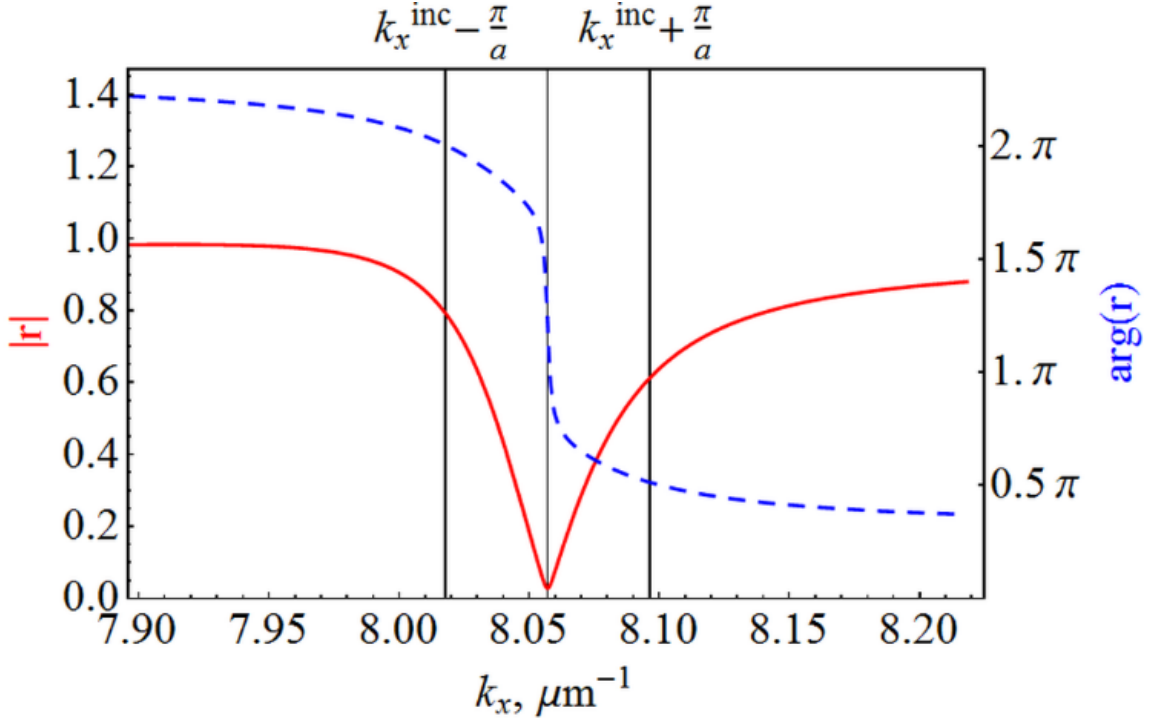

FIG. 11. The reflection coefficient $r(k_x)$, Eq. (1). Vertical lines show the values of the wavenumbers $k_x = k_x^{inc}$, $k_x = k_x^{inc} - \pi/a$, and $k_x = k_x^{inc} + \pi/a$. The magnitude and the argument of the reflection coefficient are shown by the solid red line and the dashed blue line, respectively.

Outside the geometric boundaries of the beam ($|x| > a/2$), the reflected field emerges due to interference in the plane waves.[1] Our numerical calculations show (see Fig. 9b) that in this region, the field increases with the distance from the metal surface. This behavior is a feature of a leaky wave, such as a *pS*-SPP. To prove that this field coincides with the field of the *pS*-SPP we should consider the integrand of Eq. (18) on the complex plane of $k_x$. This integrand tends to zero as $\mathrm{Im}\, k_x$ tends to infinity. Thus, integral (18) reduces to the residue at the point $k_x = k_{pS}$,

---

[1]Indeed, if $|Z_0| \gg \lambda$ (in our calculations, we use $|Z_0| = 10\lambda$), the evanescent waves in $H_i(x,0)$ nearly vanish at the metal surface. Our calculations show that in this case the value of the integral in Eq. (18) is practically independent of the upper limit $k_x^{(up)}$ for $k_x^{(up)} > k_0(\varepsilon_D)^{1/2}$. In particular, even if traces of the evanescent waves composing the beam are removed, with the computer accuracy, the reflected field would not change.

$$H_r(x,0) = \int_{-\infty}^{\infty} h_i(k_x,0) r(k_x) \exp(ik_x x) dk_x \approx 2\pi i h_i(k_{pS},0) \operatorname{Res}\left[r(k_{pS})\right] \exp(ik_{pS} x). \quad (21)$$

The field described by Eq. (21) coincides with the field of a leaky *pS*-plasmon.

In obtaining Eq. (21) we take into account contributions from residues only. To evaluate the accuracy of this approach, we have calculated integrals (15), (18) numerically. As we can see from Fig. 12, the incident field (15) yields a rather sharp beam edge (the solid line in Fig. 12) near the geometrical beam boundary $x = a/2$. As one can see from Fig. 12a, the results given by Eqs. (18) and (21) (shown by dashed and dotted lines, respectively) differ slightly for the reflected wave.

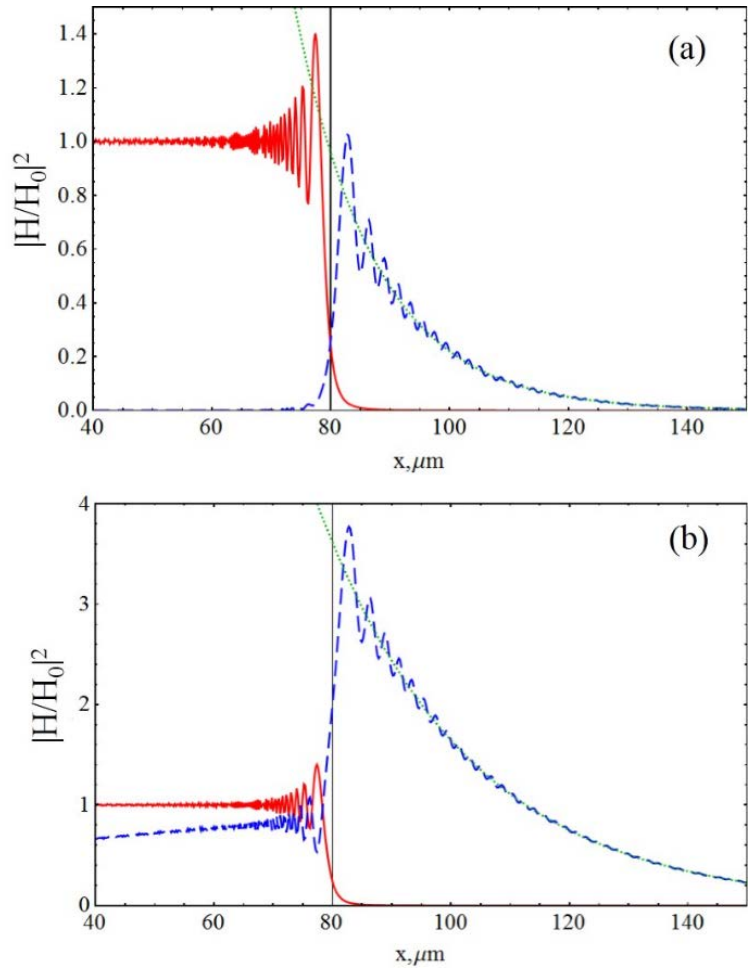

FIG. 12. The distribution of the magnetic field intensity (normalized by that in the slit) at a dielectric-metal interface, $z = 0$, upon incidence of a wave transmitted through a slit with the width of $a = 200\lambda$ situated at the distance of $10\lambda$ from the metal film; (a) actual losses and (b) losses decreased by a factor of 10. The wavelength in the free space is $\lambda = 0.8\,\mu m$. Fields of the reflected beam, Eq. (18), and an SPP, Eq. (21), are shown by blue dashed and green dotted lines, respectively. The vertical line indicates the geometrical boundary of the beam, $x = a/2$.

The interference of the plane waves (18) results in the field of a plasmon, Eq. (21), if and only if the pole of the reflection coefficient is inside the light cone. In this case, integral (18) is mainly determined by the residue of this pole. At the same time, the SPP becomes a leaky wave. We emphasize that this situation is general: if the field of the reflected beam is determined by a pole, a leaky wave is formed at the edge of this beam.

## V. CONCLUSIONS

We have shown that in the Kretschmann configuration, an SPP can only be excited if the translational invariance of the system is broken. This can be due to either surface inhomogeneity or the finite width of the light beam. In the latter case, interference of the reflected waves composing the beam results in an additional field outside the geometrical edge of the reflected beam. We demonstrate that if the angle of incidence of the beam is close to the Kretschmann angle, this additional field and the field of the SPP coincide.

Each of the plane waves composing the beam does not excite an SPP, but altogether the sum (interference) of the reflected waves results in an SPP running out of the beam. It is essential that there is an edge that violates translational invariance, resulting in a possibility of an SPP excitation.

For a leaky wave, the field propagates at an acute angle with respect to the direction of the SPP propagation. In this wave, the field increases, above the point of observation due to radiation by surface currents of segments located before the observation point. In these segments, the surface current, which decays with the propagation of the SPP along the surface, is much greater than the current at the point of observation. Since the SPP originates at the beam edge, the increase in the field amplitude is limited by the SPP amplitude at this point. Therefore, the field vanishes at large distances from the metal surface. Though the plasmon, as an eigensolution, has an infinite value, the field formed at the geometrical boundary of the beam coincides with the eigensolution in a bounded volume, and the problem of energy divergence does not arise.

Our computer simulation shows that the formation of a *pS*-SPP on the beam edge occurs even in lossless media. In this case, the minimum in the reflection coefficient does not arise. Thus, no energy is transferred to the excitation of the *pS*-SPP that emerges from the interference of reflected waves.

## ACKNOWLEDGEMENTS

The authors acknowledge the support of the Advanced Research Foundation under the Contract No. 7/004/2013-2018. A.A.L acknowledges the support of the National Science Foundation under Grant No. DMR-1312707.